\documentclass[%
reprint,
showpacs,
amsmath,amssymb,
aps,
pra,
]{revtex4-2}

\usepackage{graphicx}

\usepackage{dcolumn}
\usepackage{bm}
\usepackage{bbold}
\usepackage[export]{adjustbox}

\usepackage{mathtools} 

\DeclarePairedDelimiterX\braket[2]{\langle}{\rangle}{#1 \delimsize\vert #2}

\usepackage{array,colortbl,xcolor}

\usepackage[utf8]{inputenc}
\usepackage[T1]{fontenc}

\usepackage{mathrsfs}
\usepackage{dsfont}

\usepackage{amssymb}
\usepackage{amsmath}
\usepackage{amsfonts}
\usepackage{amsthm}

\usepackage{hyperref}
\hypersetup{pdfpagemode=UseNone}

\newcounter{rem}
\newcommand{\mc}[1]{\mathcal{#1}}

\def\>{\rangle}
\def\<{\langle}

\renewcommand{\rho}{\varrho}

\def\textbf#1{{\bf #1}}

\usepackage{xcolor}
\usepackage{soul}

\begin{document}
 
\title{The principle of majorization: application to random quantum circuits}
		 
\author{Ra\'ul O. Vallejos}
\email{vallejos@cbpf.br}
\affiliation{Centro Brasileiro de Pesquisas F\'{\i}sicas (CBPF), Rua Dr.~Xavier Sigaud 150, 
             22290-180 Rio de Janeiro, Brazil}
\author{Gabriel G. Carlo}
\email{carlo@tandar.cnea.gov.ar}
\affiliation{ Departamento de F\'{\i}sica, 
              Comisi\'on Nacional de Energ\'{\i}a 
              At\'omica, 
              Avenida del Libertador 8250, 
             (1429) Buenos Aires, Argentina}

\author{Fernando de Melo}
\email{fmelo@cbpf.br}
\affiliation{Centro Brasileiro de Pesquisas F\'{\i}sicas (CBPF), Rua Dr.~Xavier Sigaud 150, 
             22290-180 Rio de Janeiro, Brazil}

\date{\today}

\begin{abstract}
We test the principle of majorization [J. I. Latorre and M. A. Mart\'{i}n-Delgado, 
Phys. Rev. A {\bf 66}, 022305 (2002)] in random circuits. 
Three classes of circuits were considered: 
(i) universal, (ii)  classically simulatable, and (iii) neither universal nor classically simulatable.
The studied families are: 
\{CNOT, H, T\}, \{CNOT, H, NOT\},  \{CNOT, H, S\} (Clifford), matchgates, and IQP  (instantaneous quantum polynomial-time).
We verified that all the families of circuits satisfy on average the principle of decreasing majorization.
In most cases the asymptotic state (number of gates $\to \infty$) behaves like a random vector.
However, clear differences appear in the {\em fluctuations} of the Lorenz curves associated to asymptotic states. The fluctuations of the Lorenz curves discriminate between universal and
non-universal classes of random quantum circuits, and they also detect the complexity of some non-universal but not classically efficiently simulatable quantum random circuits.
We conclude that majorization can be used as a indicator of complexity of quantum dynamics, as an alternative to, e.g., 
entanglement spectrum  and out-of-time-order correlators (OTOCs).
\end{abstract}


\maketitle

\section{Introduction}
\label{sec:intro}

Majorization is a mathematical concept that allows one to decide whether a probability distribution is more disordered/spread than another. 
Offering an answer to such a fundamental question, the notion of majorization has found applications in many fields of social and natural sciences~\cite{marshall09}. 
The first known use of the majorization notion dates back to the beginning of the twentieth century, 
and it is due to the economist Max Otto Lorenz who used it as a measure for wealth concentration~\cite{lorenz}.

Within physics and information theory, the concept of majorization precedes that of entropy. 
If a probability distribution associated to a random variable $X$ majorizes the probability distribution of a random variable $Y$ (see definition in Sec.~\ref{sec:circuits}), then the (Shannon) entropy of $Y$ is bigger than that of $X$. However, the reverse statement is not necessarily true: if the entropy of $Y$ is bigger than that of $X$, it is not necessarily the case that $X$ majorizes $Y$. 
The majorization relation is thus a finer criteria to compare the spread of probability distributions than the entropy comparison~\cite{ruch75}. 
With that in mind, Ruch and collaborators proposed a stronger version of the second law of thermodynamics, known as the \emph{principle of increasing mixing character}~\cite{ruch75,ruch76,ruch78}: the time development of a statistical ensemble of isolated systems proceeds in such a way that the probability distributions at earlier times majorize those at later times. 

As expected, this intuition remains true in the quantum arena, 
with the majorization concept also playing  an important role in quantum and nano thermodynamics 
(see, e.g., \cite{horodecki03,allah04,horodecki13,renes14,egloff15}). 
Within quantum information theory, Nielsen \& Vidal's theorem~\cite{nielsen01} 
asserts that majorization determines the possible transformations between bipartite entangled states by means of local operations and classical communication. 
One recent example related to quantum computation: in the context of boson sampling~\cite{aaronson13}, 
Chin and Huh explained quantitatively that the complexity of the computation relates to the majorization ordering of the input and output particle-distribution vectors~\cite{chin17}.

While generic dynamics proceeds in the direction of growing disorder, Or\'us et al noticed that in many quantum algorithms the opposite is true: 
in the course of the computation the state of the system is step-by-step majorized, with the final result being maximally ordered \cite{orus02,latorre02}. 
Later on, by carrying a systematic analysis of a variety of quantum algorithms, 
these authors concluded that  fast and efficient algorithms (quantum Fourier transform, Grover's algorithm, the algorithm for the hidden affine function problem, and others) 
obey a {\em majorization principle} (MP), i.e., satisfy step-by-step majorization. 
Supporting this conjecture, they gave examples of some quantum algorithms not showing any computational speed-up which violate the MP~\cite{orus04}.
Further evidence supporting the MP came from adiabatic algorithms~\cite{wei06}. 
Recently step-by-step majorization for the Fourier Transform was observed experimentally in photonic circuits \cite{flamini16}. 
A detailed analysis of the Grover algorithm can be found in \cite{garcia18}.

In recent years random  quantum circuits have grown in importance. Besides various applications in quantum information and communication~\cite{hayden2004,brown2015, brandao2016}, random quantum circuits are becoming the test-bed for the so called quantum advantage. Even within the noisy intermediate scale quantum (NISQ) era, when quantum computers are composed of around hundred qubits and are still not amendable to error correction, sampling from random quantum circuits was proved a hard task for classical simulations~\cite{harrow2017,bouland2019}, and as such it is a clear demonstration of quantum advantage. Using random quantum circuits of 53 qubits,  such quantum advantage was experimentally reported in Ref.~\cite{arute2019}. 

To determine properties of quantum random circuits that help to characterize their ``complexity'' is thus an important task. Recently, in a series of articles~\cite{chamon14, shaffer14, yang17}, 
it was established an apparent connection between the entanglement spectrum statistics of 
the output of a random circuit
and its complexity -- as defined by the universality of the gate set determining ensemble. 
More concretely, let $\mc{G}$ be a set of quantum gates, and take an ensemble of $N$ qubit quantum circuits formed by uniformly sampling from  $\mc{G}$. Initiating the circuit with a random product state,  the authors observed that the level spacing statistics for the entanglement spectrum (the Schmidt values of the balanced partition) at the end of the circuit follows the random matrix theory (RMT) prediction if $\mc{G}$ forms a universal set of gates,  and a Poissonian distribution in the opposite case \cite{shaffer14}. 
The authors go one step further and suggest a connection
between the impossibility of returning from a stationary entangled state to a separable state 
(via a Metropolis like algorithm) and the complexity of the random circuit: 
universal set of gates would lead to an effectively irreversible entanglement dynamics 
(when the sequence of gates leading to the stationary state is forgotten). 

Another characteristic of complex quantum dynamics is that local information spreads quickly over the whole system.
Widely used indicators of this process are the out-of-time ordered correlators (OTOCs):
Exponentially fast changes of OTOCs have been taken as a sign of ``quantum chaos'', i.e., quantum complexity
\cite{larkin69,maldacena16,swingle18}.

It is the aim of the present work to analyze the majorization principle in different ensembles 
of random quantum circuits. 
Here we show that not only the majorization principle holds (on average) for random circuits,  
but the fluctuations of cumulant vectors (see below) do seem to correctly identify complex computation.
In addition, we show that non-universal but classically non-simulatable quantum circuits 
might also lead to a RMT entanglement spectrum. 
Our calculations demonstrate that this intermediate class of circuits can be identified 
by the majorization criteria.

\section{Majorization \& random circuits} 
\label{sec:circuits}

In the following we define majorization, 
succinctly describe the various types of quantum random circuits used,
and touch upon some related questions.

\subsection{Majorization}

Let $x$ and $y$ be real vectors of length $N$. It is said that 
$x$ is majorized by $y$ (or $y$ majorizes $x$), denoted by $x \prec y$, 
if, for all $k<N$, 
\begin{equation}
\sum_{i=1}^k x_i^\downarrow \leq \sum_{i=1}^k y_i^\downarrow \, ,
\label{maj1}
\end{equation}
and
\begin{equation}
\sum_{i=1}^N x_i^\downarrow   =   \sum_{i=1}^N y_i^\downarrow \, ,
\label{maj2}
\end{equation}
where $x^\downarrow$ means that the components of $x$ have been arranged in nonincreasing order.
The partial sums in the equations above will be called {\em cumulants}. 
The $k$-th cumulant of $x$ will be denoted $F_x(k)$. 
Here we will always use majorization for comparing probability vectors, i.e., real vectors of non-negative components 
and normalized to unity: $x_i \ge 0, \forall \,\, 1\le i \le N$  and $F(N)=1$.  

A very useful way of visualizing the majorization (partial) ordering $x \prec y$ is by plotting 
the {\em Lorenz curves} $F_x(k)$ and $F_y(k)$ vs $k/N$~\cite{lorenz}. 
Then, one has that $x \prec y$ iff the Lorenz curve for $y$ is above the curve for $x$ for all values of $k/N$.
Some examples of Lorenz curves can be seen in Fig.~\ref{fig1}. 
Note that there exist vectors $x$ and $y$ for which neither $x$ majorizes $y$, nor $y$ majorizes $x$.

\subsection{The random circuits}
\label{sec:circ}

We consider unitary quantum circuits with $n$ qubit lines ($n=8$ in our numerical calculations). 
The total system dimension, and the dimension of the probability vectors, is thus $N=2^n$. 
The system evolution through the quantum circuit is given in unit time steps. 
Fixed a gate set $\mc{G}$, at each time step a gate from $\mc{G}$ is chosen and it is applied to a selected set of qubits. 
In a quantum \emph{random} circuit the choice of gate to be applied at a given time, 
and the selection of qubits to which this gate is going to be applied, 
are both probabilistic -- as defined by preset measures. 
For each step we evaluate the state in the computational basis, the associated probabilities, 
and the cumulants $F(k)$, $k=1,\ldots,N$. This is repeated for a number $N_s$ of time steps.

We always use completely factorized initial states, i.e., 
$| \psi \rangle = | \psi_1 \rangle \otimes \ldots \otimes | \psi_n \rangle.$
Each $| \psi_i \rangle$ may be random or a fixed state. 
In the first case, we chose each factor state independently 
and uniformly distributed on the Bloch sphere (Haar measure for vectors).
Diagonal-gate circuits (IQP) use always $| 0 \rangle^{\otimes n}$ as input.
In some cases we tested both kinds of initial states for a given type of circuit.

Starting from a separable state, the successive application of gates eventually leads to highly entangled states 
showing statistical properties typical of random vectors. 
This is clearly true for the universal sets, e.g., {\em G3} (see below). 
Likewise, most families of circuits considered here show a similar behavior. 
In particular, the Lorenz curves associated to asymptotic (large number of gates) states 
are very close to those corresponding to random complex vectors.
We denote the ensemble of $n$-qubit states, with the induced Haar measure, as {\em Haar-n}.

When speaking of classical simulatability of quantum circuits one can distinguish between two main notions:
strong (all output probabilities are calculated) or weak (only a sample of the output probability is required).
It is said that a simulation by classical means is efficient if it runs in polynomial time in the input size.
The efficiency of a classical simulation may depend on the input type (arbitrary product state or computational basis state), 
and on how many output lines are measured (single-line or multiline measurement).
Other ingredients like intermediate measurements, post-selection, etc., will not be considered.
We have given a very compact definition of (efficient) classical simulatability. 
Thorough definitions may be found in the references:  
\cite{gottesman98,clark08,nest10,jozsa14,koh17} (Clifford circuits),
\cite{bremner11,xi13,fujii17} (IPQ/diagonal circuits),
\cite{valiant02,jozsa08,brod14} (matchgates).

We employed seven classes of circuits: {\em G1, G2, G3, MG, D2, D3, Dn}.
They are described in the following three sections.

\subsubsection{Circuits constructed from a few generators}
We considered the three sets used in Ref.~\cite{shaffer14}, i.e.,
{\em G1}=\{CNOT, H, NOT\}, 
{\em G2}=\{CNOT, H, S\}, and 
{\em G3}=\{CNOT, H, T\},
where CNOT is the controlled-NOT gate, H stands for Hadamard, and S and T are 
$\pi/4$ and $\pi/8$ phase gates, respectively.
The set {\em G3} is universal and thus approximates the full unitary group $U(N)$ to arbritary precision.
Both sets {\em G1} and {\em G2} contain only Clifford gates, thus are nonuniversal and classically simulatable 
(in the setting of the Gottesman-Knill theorem \cite{gottesman98,jozsa14}).
The circuits constructed from {\em G2} generate the Clifford group \cite{gottesman98}.
The gates in {\em G1} generate a subgroup of Clifford, 
then the simulatability of {\em G2} implies the simulatability of {\em G1} (in the same settings).

The probability for a gate to be selected at given time is  always 1/3 for the four families above.
We also choose with equal probability the qubits or pairs of qubits to which a selected gate is applied.

\subsubsection{Matchgate circuits (MG)}
Matchgates are two-qubit gates formed from two one-qubit gates A and B with the same determinant:
A acts on the even parity subspace (spanned by $|00\rangle$ and $|11\rangle$)
and B acts on the odd parity subspace (spanned by $|01\rangle$ and $|10\rangle$).
A and B are randomly chosen according the Haar measure in the unitary group $U(2)$. 
All pairs of qubits are equiprobable.
Circuits of matchgates acting on nearest-neighbor lines only  are classically simulatable; however,
if the nearest-neighbor restriction is lift, the resulting circuits are universal for quantum computation 
\cite{jozsa08,brod14}.

\subsubsection{Diagonal-gate circuits (D2, D3, Dn)}
\label{sec:diag}
These circuits are made up from gates which are diagonal in the computational (Z) basis.
The initial state is set to $| 0 \rangle^{\otimes n}$ and 
Hadamard gates are placed at the beginning and ending of each line \cite{bremner11}.
This class of circuits is also called IQP (instantaneous quantum polynomial-time):
As diagonal gates commute, they can be applied simultaneously and thus there is not a natural
time ordering of gates for a given circuit.

Diagonal circuits cannot perform universal computation, however, in general, they are not
classically simulatable \cite{shepherd09,bremner11}.

We shall deal with a particular subclass of diagonal circuits: the $r$-qubit phase-random circuits  
\cite{nakata14a}. 
In this case, each gate acts upon $r$ qubits and has the form
\begin{equation}
W_r={\rm diag} \{ e^{i \phi_1},e^{i \phi_2},\ldots,e^{i \phi_{2^r}} \} \; ,
\end{equation}
with the $\phi$'s independent random uniform in $[0, 2 \pi)$.
The gates are applied on all combinations of $r$ (out of $n$) qubits, the ordering being random.
Here we use only three values for $r$, namely $r=2,3,n$.
For $r=2$ the circuits contain $n(n-1)/2$ gates, while for $r=n$ the circuits consist of only one random gate.
So, our diagonal circuits have a well defined number of gates.
On the contrary, {\em G1, G2, G3} and {\em MG} may have an arbitrary length. 

In the case $r=2$ we also consider the situation of gates acting on nearest-neighbor qubits. 
Assuming a ring topology, these circuits contain only $n$ gates, and they are classically simulatable \cite{fujii17}.

\subsection{Abbreviations}
As we are considering various types of circuits, connectivity, and initial states, 
for the sake of compactness, we shall employ abbreviations to designate the different
possibilities. 
To a triplet \{circuit, connectivity, initial state\} we shall associate the word {\em A-B-C}.
{\em A} stands for the type of circuit: {\em G1, G2, G3, MG, D2, D3, Dn};
{\em B} may equal {\em nn} (nearest neighbor), 
{\em rn} (random neighbors), 
or {\em all} (all combinations of $r$ qubits --only for diagonal circuits);
and {\em C} may be {\em rs} (random initial state) or {\em 0} ($| 0 \rangle^{\otimes n}$ as input).

For instance, {\em MG-rn-rs}, denotes a circuit composed from matchgates acting on arbitrary neighbors,
with random initial state. 
Analogously, {\em D3-all-0} denotes a diagonal 3-qubit random phase circuit.

Before proceeding with the numerical calculations, we note that, for our purposes, 
the various families of circuits to be analyzed can be split into three categories: 
(i) universal, (ii)  classically simulatable, and (iii) neither universal nor classically simulatable.
We shall try to correlate the different features in our calculations with these three classes.

\section{Numerical calculations} 
\label{sec:num}

\begin{center}
\begin{figure}[t!]
\includegraphics[width=1.0\linewidth,angle=0]{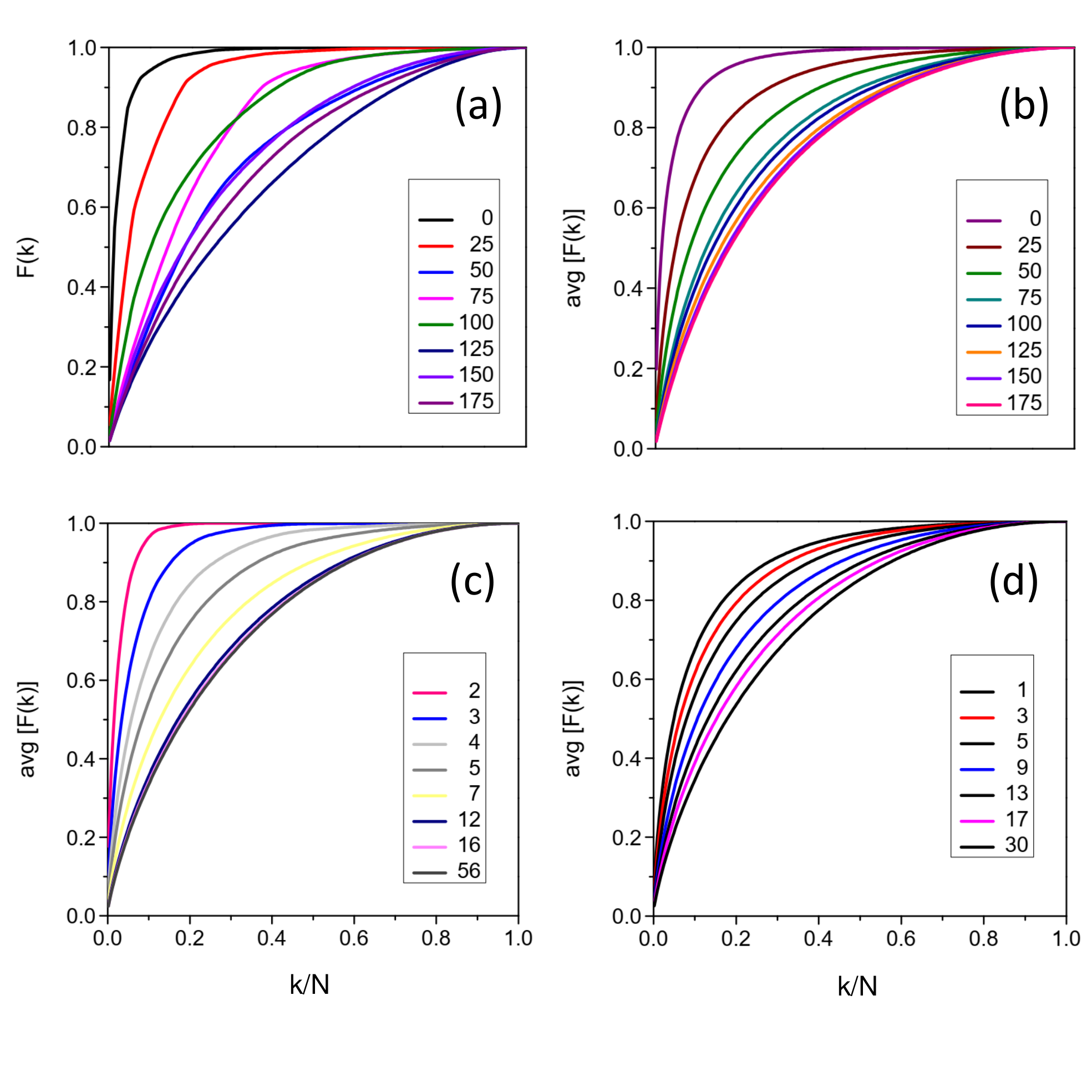}
\caption{ 
(a) Lorenz curves for a single realization of a circuit belonging
to the family {\em G1}=\{CNOT,H,NOT\}. 
The topmost curve corresponds to the initial state. 
The other curves were calculated after the action of 25/50/75/100/125/150/175 gates
(the color scheme for the different numbers of gates is shown in the inset).
(b) Lorentz curves for the family {\em G1} but now averaged over 500 realizations. 
The numbers of gates are the same than in (a).
(c) Same as (b) but for diagonal  {\em D3-all-0} circuits. 
The curves were calculated after the application of 2/3/4/5/7/12/16/56 gates.
(d) Same as (b) but for matchgates {\em MG-rn-rs} circuits.
Number of gates are 1/3/5/9/13/17/30.
All circuits have 8 qubit lines.
}
\label{fig1}
\end{figure}
\end{center}

We start by giving an example of Lorenz curves for a particular realization of a circuit {\em G1}
and a random initial state. 
In Fig.~\ref{fig1}(a) we plot a set of Lorenz curves for different times. 
We see that, as time grows, there is an overall tendency to decreasing majorization.
For example, if $x_t$ denotes the probability vector at time $t$, then we have  $x_0 \succ x_{25} \succ x_t$, 
for $t= 50,75,100,\ldots$.
However, there are cases of increasing majorization (curve swapping, e.g., $x_{125} \prec x_{150}$) 
as well as cases of no definite majorization order (curve crossing, e.g., $x_{75}$ and $x_{100}$).
We have observed analogous behavior for other realizations of {\em G1}, and also for circuits belonging
to other families. 

Not surprisingly, crossings and swappings vanish by averaging over realizations and/or initial conditions, 
and then majorization order appears. 
This can be clearly seen in Fig.~\ref{fig1} [panels (b)-(d)], where we exhibit Lorenz curves for selected times,
but now averaged over 500 circuits/initial states. 
Results for families {\em G1-rn-rs}, {\em D3-all-0}, and {\em MG-rn-rs} are qualitatively similar, 
all showing decreasing majorization as the number of gates is increased.
%
%
Though the rate of change of the Lorenz curves depends of the type of circuit, 
the asymptotic curves are almost identical.
From here on, we focus on the asymptotic Lorenz curves -- asymptotic meaning large number of gates for
circuits {\em G1, G2, G3, MG} or the maximum number of gates for {\em D2, D3, Dn}.

\begin{center}
\begin{figure}[t!]
\includegraphics[width=1.0\linewidth,angle=0]{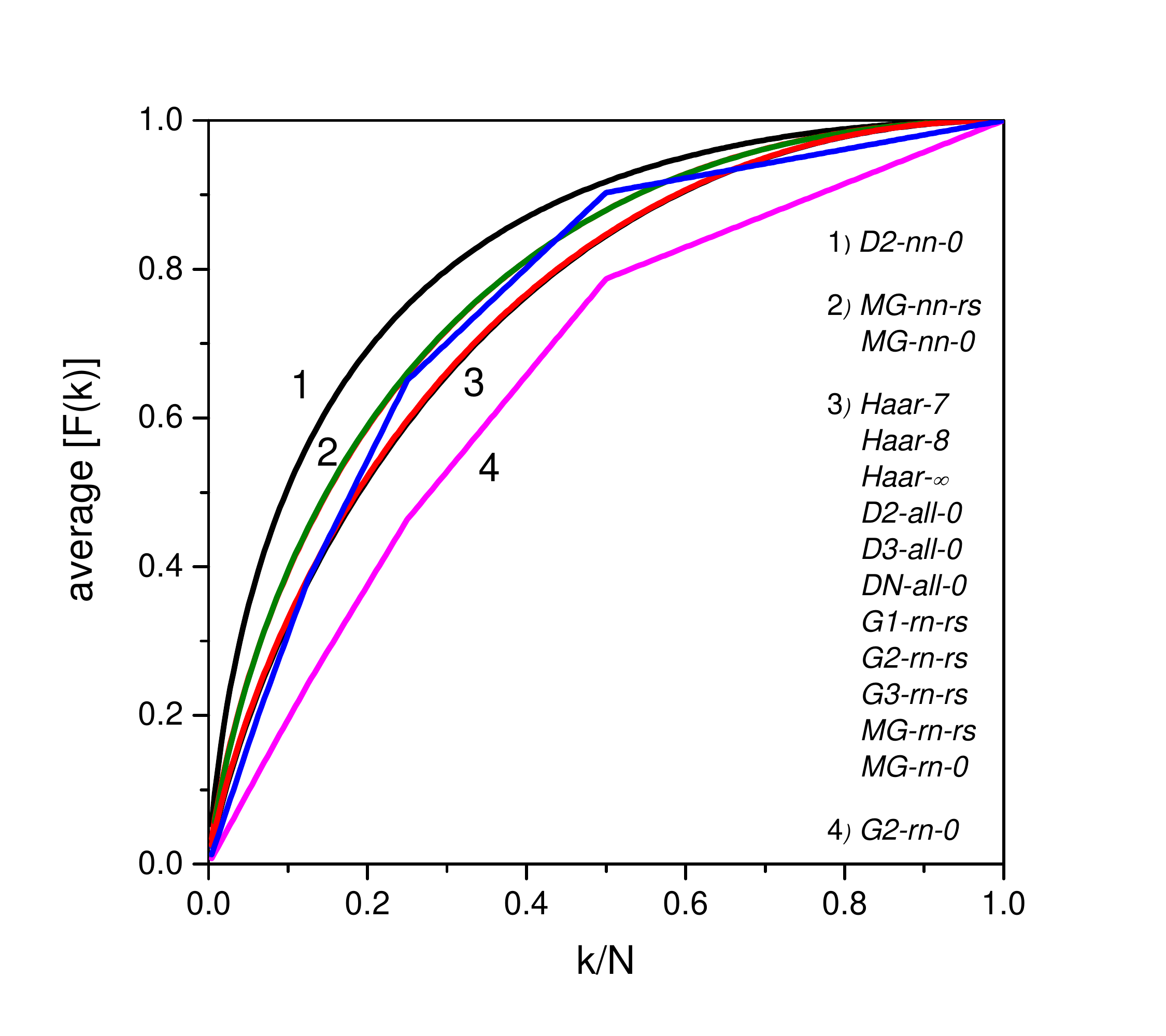}
\caption{
Average Lorenz curves for asymptotic states generated by various families of random circuits of 8 qbits. 
For circuits {\em G1, G2, G3, MG} we used 500 gates to ensure convergence.
For the diagonal ones, i.e., {\em D2, D3, Dn}, the maximum number of gates was used (see Sec.~\ref{sec:diag}).
Inset: circuit names ordered to correspond to curve ordering (decreasing majorization), 
except for {\em G1-rn-0} (broken blue line). }
\label{fig:avg}
\end{figure}
\end{center}

Figure~\ref{fig:avg} depicts averages of the asymptotic Lorenz curves for all classes of circuits considered.
Most curves coincide (within visual resolution) with that corresponding to random vectors of 8 qubits 
({\em Haar-8}).  

We have verified numerically that the Lorenz curves associated with random vectors tend to a limiting curve, Haar-$\infty$, 
as the dimension increases. In Fig.~\ref{fig:avg} we see that this limit is reached already for 7 qubits. 

%
%

Some comments about Fig.~\ref{fig:avg} are in place.
We have also plotted {\em Haar-7} because
matchgates preserve the symmetry of the initial state,
then, if we start from $|0\rangle^{\otimes n}$, the final states will live in the positive parity subspace.
This subspace has dimensionality $2^n/2$, i.e., the final state-vector has $2^n/2$ null components.  
We discarded these null components, thus considering state-vectors of half the dimension of the initial state.
Because of this, results for {\em MG-rnd-0} and {\em MG-nn-0} must be compared with {\em Haar-7}.

The circuits {\em D2-nn-0} and {\em G2-rn-0} produce average Lorenz curves which deviate the most from
the universal, random vector, behavior (presumably {\em D2-nn-0} has too few gates for scrambling the initial state).Note that both circuits are classically simulatable under fairly undemanding settings \cite{fujii17,gottesman98}. {\em G1} and  {\em G2} (Clifford) circuits acting on the state  $|0\rangle^{\otimes n}$ produce states with highly 
degenerate amplitudes \cite{dehaene03,nest10}. Then, the corresponding Lorenz curves are piece-wise linear.

\begin{center}
\begin{figure}[t!]
\includegraphics[width=1.0\linewidth,angle=0]{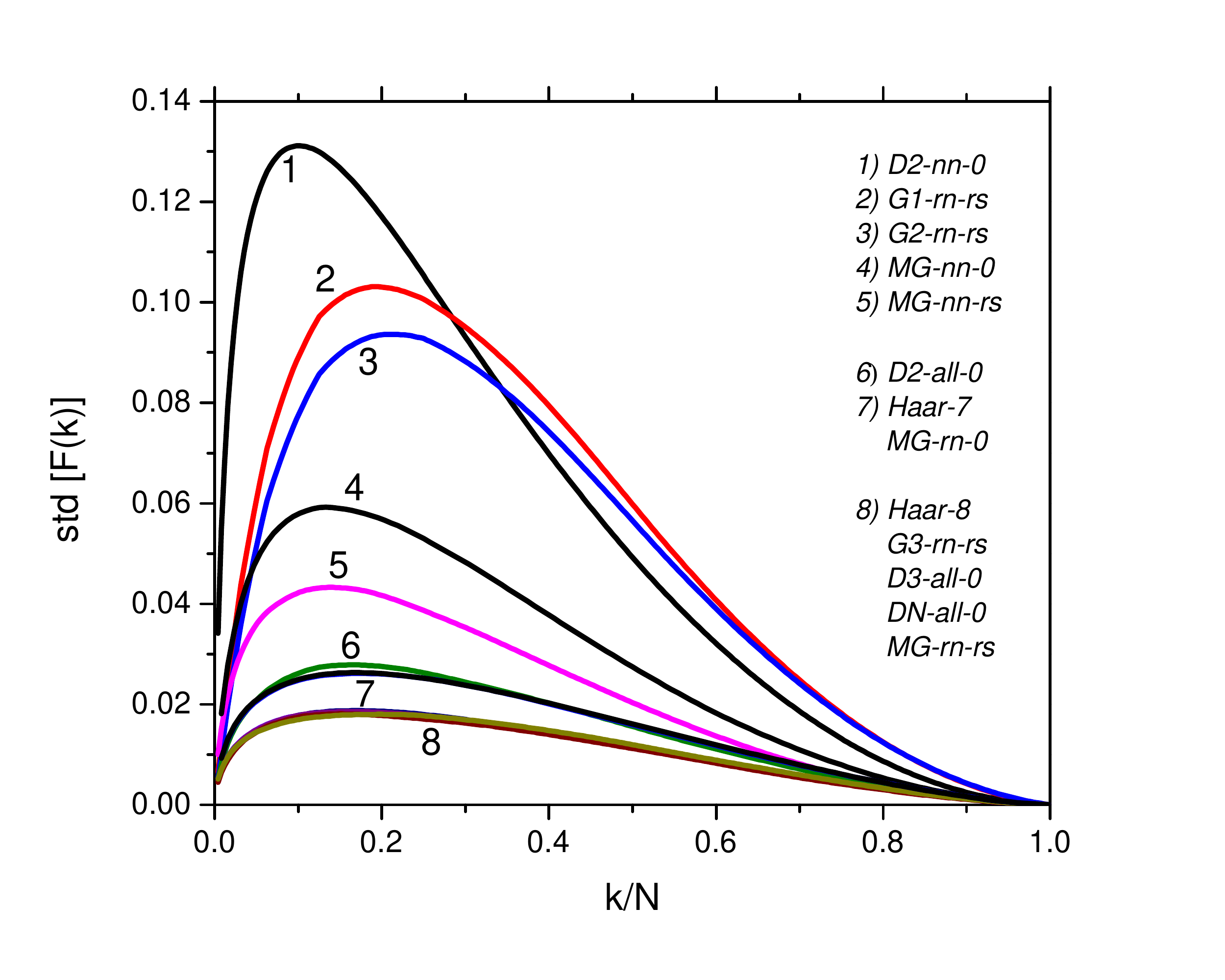}
\caption{ 
Fluctuations (standard deviation) of Lorenz curves for asymptotic states generated by various families of 
random circuits of 8 qubits.
For each family we considered 5000 samples of 500 gates, except for diagonal circuits which have fixed lengths.
Inset: circuit names. Name ordering corresponds to curve ordering. 
The families {\em G1-rn-0} and {\em G2-rn-0} show much larger fluctuations;
their Lorenz curves are piecewise quadratic (not shown).}
\label{fig:sig}
\end{figure}
\end{center}

Having seen that average Lorenz curves fail to differentiate universal from nonuniversal families, 
we decided to look at the {\em fluctuations} of the Lorenz curves, i.e., the 
fluctuation of each cumulant $F(k)$ for each class of circuits, averaged over many realizations.
The result of this calculation is shown in Fig.~\ref{fig:sig}.

%
%

Here we observe a larger spread of the Lorenz curves, with a partial removal of the
coalescence of Fig.~\ref{fig:sig}: {\em G1} and {\em G2} (both nonuniversal)
have separated off the main group, 
exhibiting larger fluctuations.
The smallest, universal fluctuations correspond to the circuits 
{\em G3-rn-rs, MG-rn-rs, DN-all-0, D3-all-0} (coincide with {\em Haar-8}),
and {\em MG-rn-0} (coincides with {\em Haar-7}).

The family {\em D2-all-0} stands as a special case because, in spite of showing
very small fluctuations does not match the universal prediction for circuits without
symmetries, i.e., {\em Haar-8}. We return to this point below.

We have verified numerically that the fluctuations corresponding to random vectors 
decrease as the number of qubits grows (cf. Fig.~\ref{fig:sig}). 
This raises the questions: What is the associated scaling law? 
Do the fluctuations of the various families obey different laws?
Answering these questions, though a source of potentially valuable information, 
lays outside of the scope of the present paper.

\subsection{Entanglement spectra} 
\label{sec:spec}

Following \cite{shaffer14} we considered a 50-50 bipartition and 
calculated the spectra of the asymptotic reduced matrices.
Let us denote the eigenvalues $\lambda_1,\lambda_2,\ldots,\lambda_{N/2}$, in decreasing order.
Use the gaps $\epsilon_i=\lambda_{i+1}-\lambda_i$ to form the quotients $r_i=\epsilon_{i+1}/\epsilon_i$.
Then construct the histogram $P(r)$ by sampling within each class of circuit.
If the distribution of gaps follows a Wigner-Dyson law (characteristic of Gaussian random matrix ensembles \cite{mehta04}),
the theoretical prediction for $P(r)$, in the limit of large matrices, is \cite{atas13}
\begin{equation}
P_{\rm WD}(r)=\frac{1}{Z} \frac{\left(r+r^2\right)^\beta} {\left(1+r+r^2\right)^{1+3 \beta/2}} \;,
\label{eq:WD}
\end{equation}
where $Z = \frac{8}{27}$ for the Gaussian Orthogonal Ensemble (GOE) with $\beta = 1$, 
and $Z = \frac{4}{81} \frac{\pi}{\sqrt{3}}$
for the Gaussian Unitary Ensemble (GUE) with $\beta= 2$.

On the other side, if the spectrum is uncorrelated, then the corresponding statistics if Poissonian:
\begin{equation}
P_{\rm Poisson}(r)=\frac{1} {\left(1+r \right)^2} \;.
\label{eq:Poisson}
\end{equation}

Given that the circuits considered here do no exhibit time-reversal symmetry, their spectral statistics
should follow the predictions of GUE -- provided they exhibit complex dynamics. 
In fact, it was shown in \cite{shaffer14} that $P(r)$ for the universal family {\em G3}  
agrees with GUE, i.e., Eq.~(\ref{eq:WD}) with $\beta= 2$. 
On the other side, the families {\em G1} and {\em G2} showed Poisson-like fluctuations.

Our matrices are not large enough to use the asymptotic expressions above. 
So, in order to account for finite dimensions effects, we resorted to numerical calculations.
In the case of GUE statistics, instead of employing Eq.\eqref{eq:WD}, we used the spectra of reduced density matrices obtained via partial
trace of Haar random vectors of $n$ qubits. The results obtained from these spectra will be labeled Haar-$n$. 
In the other side, we generated Poissonian spectra simply by choosing $N/2$ levels $\in (0,1)$ 
independently and random-uniformly.

%
\begin{center}
\begin{figure}[t!]
\includegraphics[width=1.1\linewidth,angle=0]{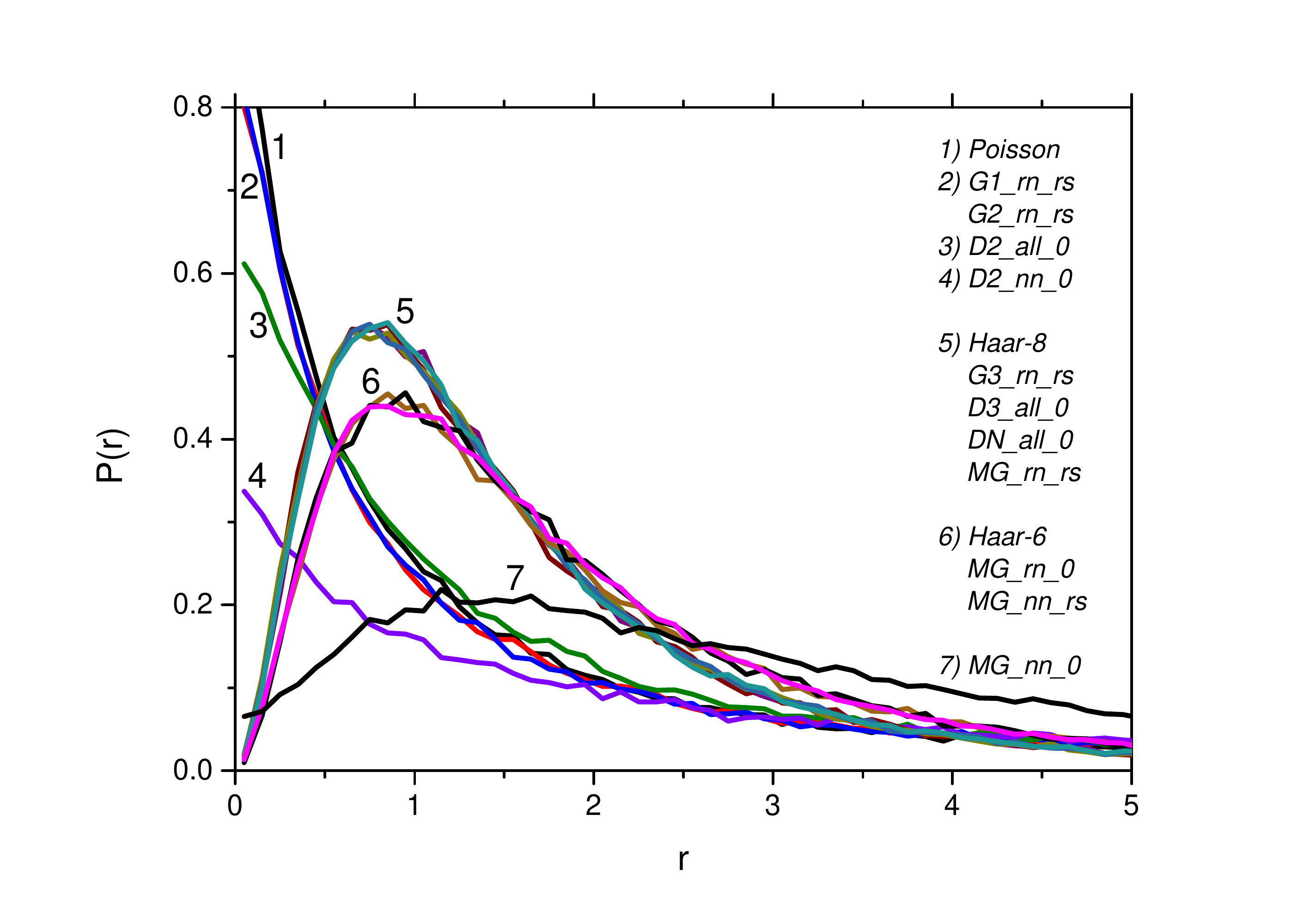}
\caption{ 
Entanglement spectra for asymptotic states generated by various families of random circuits.
Shown are averages of 5000 circuits of 500 gates.
Diagonal circuits have fixed lengths. All circuits have 8 qubits. 
 }
\label{fig:spec}
\end{figure}
\end{center}
%

We extend the spectral calculations of $P(r)$ in \cite{shaffer14} 
(restricted to {\em G1}, {\em G2} and {\em G3}) to matchgates and diagonal circuits.
In Fig.~\ref{fig:spec} we show histograms representing $P(r)$ for all the circuits.
We observe that they can essentially be divided into two groups:
(a) Poisson-like, i.e., decreasing probability distributions
({\em G1-rn-rs, G2-rn-rs,  D2-all-0, D2-nn-0}\/), and (b) GUE-like, i.e., peaked histograms.
The latter category includes the circuits  {\em G3-rn-rs, D3-all-0,  DN-all-0, MG-rn-rs\/}, 
which show excellent agreement with the GUE prediction ({\em Haar-8}), and 
{\em MG-rn-0} which coincides with {\em Haar-6\/}
(this was to be expected because {\em MG-rn-0} is universal but has
parity symmetry~\footnote{The final states of the circuits {\em MG-rn-0} possess parity symmetry. 
	It is easy to show that the corresponding reduced density matrices also are parity-symmetric. 
	Because of this, their entanglement spectra consist of two uncorrelated sequences, each one having a definite parity. 
	Thus, we separated the ensemble of spectra into two symmetry classes, and then calculated two sets of histograms. 
	Both set of histograms are consistent with the GUE distribution. 
	Some considerations apply to {\em MG-nn-0}.
	In both cases we considered only the even-symmetry spectra 
	(odd spectra produce statistically identical histograms).}). 

The spectra corresponding to the families {\em G1-rn-0} and {\em G2-rn-0} consist of several 
multiply degenerated levels. Thus, the associated $P(r)$ are singular 
(nor Poisson- neither GUE-like) and were not plotted.

In the previous sections we have limited ourselves to presenting the results of our calculations, 
reserving the discussion thereof for the next, concluding section.

\section{Discussion \& Final remarks} 
\label{sec:conclusions}		

The purpose of this paper was to characterize the complexity of quantum random circuits using majorization criteria.
The authors of \cite{shaffer14} had showed that the entanglement spectrum could be used to distinguish universal
from non-universal families of circuits, using as examples the families {\em G1, G2, G3\/}.
Here we extended the analysis of \cite{shaffer14} in two directions. 
First, we considered additional families of circuits, i.e., those constructed either from matchgates or diagonal gates.

Second, we inquired about the complexity (according to majorization criteria or entanglement spectrum) 
of the families that, in spite of being non-universal, cannot be efficiently simulated in a classical computer.
We found that some families of diagonal circuits (non-universal) must be classified as complex according to
the above mentioned criteria. 

After verifying that all circuit families satisfy the principle of decreasing majorization (on average), 
we focused on the Lorenz curves of the output states.
Our results, displayed in Figs.~\ref{fig:avg}, \ref{fig:sig}, \ref{fig:spec}, are summarized in Table~\ref{tb:qrcircuits}.
The last two columns exhibit the agreement, for all the circuits, between both complexity indicators considered.
Also included in the table are the settings for efficient classical simulatability (fourth column).
The third column displays the settings for which a given family is known to be {\em not} simulatable.
Altogether we have considered eight possible settings, corresponding to three binary choices: 
weak vs strong simulation, single- or multi-line tasks, and the input being either $\left|0 \right \rangle^{\otimes n}$
or a random product state (the latter information makes part of the circuit name).
So, when we state that a family is ``always" or ``never" simulatable, 
we mean as far as the eight above-mentioned settings are concerned. 

\begin{table*}
	\begin{tabular}{|l|c|c|c|c|c|c|}
\hline
Circuit name &  UNIV? & Non classically simulatable?  & Classically simulatable?   & Ave-H? & Fluc-H? & Spec-RMT? \\  \hline
 {\em G3-rn-rs} & Yes & Always                        & Never                           & Yes & Yes & Yes \\ 
 {\em G2-rn-rs} & No  & OUT(MANY) WEAK \cite{jozsa14} & OUT(1) STRONG \cite{clark08}    & Yes & No  & No \\  
 {\em G1-rn-rs} & No  &  (?)                          & OUT(1) STRONG                   & Yes & No  & No \\                                
 {\em G2-rn-0 } & No  & Never                         & OUT(MANY) STRONG \cite{clark08} & No  & No  & No \\   \hline
 {\em MG-rn-rs} & Yes & Always        & Never                                             & Yes & Yes & Yes \\
 {\em MG-rn-0}  & Yes & Always        & Never                                             & Yes & Yes & Yes \\
 {\em MG-nn-rs} & No  &  (?)          & OUT(1) STRONG \cite{jozsa08,brod14}               & No  & No  & No$^\ast$ \\
 {\em MG-nn-0}  & No  & Never         & OUT(MANY) STRONG \cite{valiant02,terhal02,brod16} & No  & No  & No         \\  \hline
 {\em DN-all-0} & No  & OUT(MANY) WEAK \cite{bremner11}  & OUT(1) WEAK \cite{bremner11}  & Yes & Yes & Yes \\   
 {\em D3-all-0} & No  & OUT(MANY) WEAK \cite{bremner11}  & OUT(1) WEAK \cite{bremner11}  & Yes & Yes & Yes \\
 {\em D2-all-0} & No  & OUT(MANY) WEAK \cite{bremner11}  & OUT(1) WEAK \cite{bremner11}  & Yes & No$^{\ast \ast}$ & No \\          
 {\em D2-nn-0}  & No  & OUT(MANY) STRONG \cite{fujii17}  & OUT(MANY) WEAK \cite{fujii17} & No  & No  & No  \\ \hline   
	\end{tabular}
	\caption{Summary of the results. 
	                 The first column lists the circuits. 
									 The second column says if the circuit family is universal or not.
								   The third and fourth columns inform about the classical simulatability (or not) 
									 and the settings of the corresponding proofs.  
									 The acronyms STRONG and WEAK refer to the tasks of calculating or sampling
									 from the output probability, respectively (see \ref{sec:circ}). 
									 OUT(1) and OUT(MANY) say whether the task is single- or multi-line. 	
                   Ave-H: average cumulants of the final states coincide with those of random vectors (Haar measure);
                   Fluc-H: same as before, but for the fluctuations of the cumulants;
                   Spec-RMT: reduced density matrix spectra well described by random matrix theory; 
                   (?): No results, to best of our knowledge;
                   No$^\ast$: RMT-like, coincides with {\em Haar-6}, but for this non-symmetric case RMT predicts Haar-8;
                   No$^{\ast \ast}$: Very close to {\em Haar-6}, however, as before, the prediction of RMT is {\em Haar-8}.
									 }
     \label{tb:qrcircuits}
\end{table*}

These settings present different degrees of difficulty for efficient classical simulation, ranging from
OUT(1)-WEAK (easiest) to OUT(MANY)-STRONG (hardest).
Circuit families that are simulatable under the hardest settings are farthest from 
performing universal quantum computation. 
Accordingly, these families are expected to deviate most from the universal behavior described by random matrices/vectors.
This correspondence is clearly exemplified by {\em G2-rn-0\/} and {\em MG-nn-0\/}, which exhibit the lowest 
complexity records  [both are OUT(MANY)-STRONG simulatable].
In spite of being simulatable under weaker conditions, {\em MG-nn-rn} and {\em D2-nn-0\/} also rank as the lowest 
complex circuits (however, these families might be simulatable in stronger settings~\footnote{The family {\em D2-nn-0} might be {\em almost} strongly simulatable. Indeed,  Fujii and Morimae \cite{fujii17} devised a simulatability proof for IPQ circuits formed by gates of the type $\exp(i \theta Z_j \otimes Z_k)$, with $(j,k)$ denoting nearest-neighbor qubit lines. They name the simulatability class of nearest-neighbor IQP ``almost strongly simulatable'', between strongly simulatable (in the exact sense) and weakly simulatable. However, strictly speaking, one can not assert that both {\em D2-nn-0} and nearest-neighbor IQP belong to the same simulatability class because {\em D2-nn-0} contains nearest-neighbor IQP.}
\footnote{Matchgates acting on nearest neighbors, having random product states as inputs, are OUT(MANY)-STRONG simulatable on the line (open boundary conditions) \cite{brod16}. Can this result be extended to the cycle (periodic boundary conditions, as used in the 	present paper)? Brod presents some geometric arguments in contrary, however, he admits that these may not be definitive: ``Curiously, the circuit [...] for periodic boundary conditions corresponds to a geometry where matchgates are universal [...], although it might just use this geometry in a very restricted manner that does not break the simulability'' \cite{brod16}. }).

We have written a clear-cut YES/NO in the last three columns of Table~\ref{tb:qrcircuits} based on the strict agreement (or not)
between our data and the predictions for random outputs. 
If we descend to the semi-quantitative level, the complexity indicators may become in conflict. 
For instance, {\em MG-nn-rs} [OUT(1)-STRONG simulatable] has indeed a GUE-like spectrum but 
intermediate-size cumulant fluctuations.
On the other side, {\em D2-all-0} [OUT(1)-WEAK simulatable] has Poisson-like spectrum, however 
its fluctuations are close to universal.
The case of {\em MG-nn-0\/} [OUT(MANY)-STRONG] is clearer: mildly GUE-like spectrum and large fluctuations. 
In the previous cases both indicators cooperate to characterize the respective circuit families as not complex.

Some unexpected behaviors, e.g., {\em MG-nn-rs} having a spectrum close to GUE, or, {\em D2-all-0} having Poisson-like
spectrum and small cumulant fluctuations, call for further studies. 
In particular one should analyze how the cumulant fluctuations decrease as the system size increases.
In this respect, it would be very useful to have analytical results for the fluctuations of Lorenz curves 
in the case of complex random vectors, which seems feasible~\footnote{Tobias Micklitz and Felipe Monteiro, private communication.}.

We verified that the {\em fluctuations} of the Lorenz curves qualify as an indicator of complexity,
producing essentially the same classification as the entanglement spectrum. Both measures, therefore, not only discriminate between universal and non-universal classes of random quantum circuits, but they also detect the complexity of some non-universal but not classically efficiently simulatable quantum random circuits.  It should be noted, however, that the fluctuations of the Lorenz curves are more easily obtained than the entanglement spectrum. The former requires simple  evaluation/measurement of the probabilities of the computational basis, while the latter requires full tomography of half the qubits and further diagonalization of the obtained reduced density matrix.

To conclude, we have introduced a criterion based on majorization (fluctuations of Lorenz curves) as an indicator of
complexity of quantum computations/dynamics. This indicator is intended to serve as an alternative or complement to 
the well known entanglement spectrum and OTOCs.

\vspace{1pc}
{\bf Acknowledgments:} 
We gladly acknowledge fruitful discussions with 
Andreas Ketterer, Daniel Brod, Eduardo Mucciolo, Ernesto Galvão, Felipe Monteiro, and Tobias Micklitz.
We are grateful to E. Mucciolo for providing us with some of the codes used in the numerical calculations.
This work is supported by the Brazilian funding agencies CNPq and CAPES, and it is
part of the Brazilian National Institute for Quantum Information (INCT-IQ).
G.G.C. gratefully acknowledges support from CONICET.


%
\end{document}